\documentstyle[preprint,aps,tighten]{revtex}
\input epsf
\epsfverbosetrue

\begin{document}

\title{Quantum time--delay in chaotic scattering: \\
       a semiclassical approach
       \thanks{Submitted to Journal of Physics A}}

\author{ R. O. Vallejos, A. M. Ozorio de Almeida}

\address{Centro Brasileiro de Pesquisas F\'{\i}sicas \\
         R. Xavier Sigaud 150, 22290-180 - Rio de Janeiro, Brazil}

\author{C. H. Lewenkopf}
\address{Instituto de F\'{\i}sica, UERJ \\
         R. S\~ao Francisco Xavier 524, 
         20559-900 - Rio de Janeiro, Brazil} 

\date{\today}

\maketitle


\begin{abstract}
We study the universal fluctuations of the Wigner-Smith time delay for
systems which exhibit chaotic dynamics in their classical limit.  We
present a new derivation of the semiclassical relation of the quantum
time delay to properties of the set of trapped periodic orbits in the
repeller.  As an application, we calculate the energy correlator in the
crossover regime between preserved and fully broken time reversal
symmetry.  We discuss the range of validity of our results and compare
them with the predictions of random matrix theories.
\end{abstract}

\draft

\narrowtext

\section{Introduction}

Over the past decade many studies were devoted to the understanding of
quantum manifestations of classical chaos.  This interest can be
explained by the fact that this subject has applications in many
different areas of physics, like properties of complex systems,
fundamental aspects of the correspondence principle, transport in
ballistic mesoscopic cavities, etc..  Most of the theoretical studies
have concentrated on spectral properties of closed systems,
accumulating a large body of numerical evidence of universality and
some analytical understanding of this fact.  Comparatively few studies
have so far been devoted to open systems, and the scattering problem
still lacks some solid fingerprints which serve to clearly distinguish
integrable from chaotic classical scattering.  For this reason, it is
desirable to study an observable which bridges the well understood
quantum aspects of closed chaotic systems and the still unclear
features of open ones.  The Wigner-Smith time delay
\cite{Wigner55,Smith60} is such an object, since it is intimately related
to the level (or resonance) density of the system and it is a genuine
scattering observable.  The present study deals with the universal
features of the time delay common to all chaotic scattering systems.
 
The concept of time delay in quantum scattering was first considered by
Eisenbud \cite{Eisenbud48} and Wigner \cite{Wigner55} in the context of
one channel scattering. Later on, Smith \cite{Smith60} extended the
previous discussions to the many channel problem by introducing the
lifetime matrix
\begin{equation}
\label{Q_ab}
Q_{ab}(E) = -i \hbar \sum_{c=1}^{\Lambda} 
              S_{ac}(E) \frac{d}{dE}S^{\dagger}_{cb}(E) ~,
\end{equation}
where $S$ is the standard scattering matrix and the sum runs over all
$\Lambda$ open channels denoted by $c$.  By averaging over the
eigenvalues of $Q$, one arrives at the so--called Wigner--Smith time
delay
\begin{equation}
\label{twsdef}
\tau(E) = -\frac{i \hbar}{\Lambda} \mbox{Tr} 
           \left[S^\dagger(E) \frac{d}{dE}S(E) \right] =
          -\frac{i \hbar}{\Lambda} \frac{d}{dE} \log \det S(E) ~,
\label{smith}
\end{equation}
which is then interpreted as the typical time spent by the particle in
the interaction region.  Even though this interpretation has some
limitations in the case of wave packet scattering
\cite{Nussenszveig72}, no difficulties arise when the incoming wave can
be considered monoenergetic, a common situation, {\sl e.g.} in
applications to mesoscopic transport phenomena \cite{Buttiker93,others}
and microwave cavity experiments \cite{Doron90,Lewenkopf92,Alt95}.

In general, one can distinguish two regimes associated with a
scattering process: a fast response (corresponding to {\sl direct}
processes) and a delayed response related to the formation of a
long--lived resonance. In the energy domain, direct processes rule the
energy--averaged behavior of $\tau(E)$. Alternatively, strong
fluctuations on the scale of the mean resonance spacing $\Delta$ are
associated to quasi--bound states, and are, in turn, intimately linked to 
the classical dynamics in the interaction region.

Our analysis deals with a specific model which illustrates very nicely
the most important properties of chaotic scattering and is well suited
to study the Wigner-Smith time delay.  Some steps in our considerations
take into account system specific properties.  However, our main
results can be easily extended to other chaotic scattering potentials.
Our model consists of an irregularly shaped cavity (denoted by $R$ in
Fig.~1) attached to a pipe (corresponding to the region $L$).  The
boundary between the pipe (or ``waveguide") and the cavity is
arbitrarily chosen at the entrance of the cavity, at $x=D$
(the region $R$ need not necessarily be a billiard).  The
quantum propagation in the direction parallel to the pipe axis is
free.  In the transversal direction there are quantized modes
$\phi_c(y)$ of energy $\epsilon_c$, defining the scattering channels
$c$.  At $x = 0$, with $D$ chosen to be sufficiently large, the wave
function $\psi(x,y;E)$ is expressed as a superposition of propagating
modes
\begin{equation}
\label{scatwave}
\psi(x,y;E)=  
\sum_{c=1}^\Lambda  
\left( a_c e^{ik_cx}  - 
       \sum_{c^\prime=1}^\Lambda  S_{cc^\prime}(E) a_{c^\prime}  
          e^{-ik_{c}x}   \right) \phi_c(y) ~,
\end{equation}
with the wavenumbers $k_c$ given by
\begin{equation}
\frac{\hbar^2}{2m} k^2_c= E-\epsilon_c  \;.
\end{equation}
From this equation it becomes evident that by choosing $|k|D\gg 1$ we
ensure
that no evanescent mode survives at $x=0$ and Eq.~(\ref{scatwave}) is
valid.  All the information about the scattering process is contained
in the energy dependent scattering matrix $S(E)$.
  
For chaotic systems, we derive a formula which shows how to calculate
the quantum time delay using information about the classical orbits
trapped inside the system.  We further explore this formula to compute
time delay correlators, which show universal features.  The universal
correlators typically scale with quantities like the average dwell time
inside the system, $\tau_{\mbox{\scriptsize dwell}}$, and the mean
resonance spacing $\Delta$ (also given in terms of the Heisenberg time
$\Delta = 2\pi \hbar/\tau_{\mbox{\scriptsize H}}$).
We show that the universal curves obtained in the semiclassical theory
are in good agreement with the statistical theory of random matrices
when $\tau_{\mbox{\scriptsize H}}/\tau_{\mbox{\scriptsize dwell}} \gg
1$.

This paper is structured in the following manner.  In Section II, a
novel derivation of the quantum time delay in terms of the underlying
classical phase space of the repeller is presented.  Some applications
of this formula are explored in Section III.  The comparison with
random matrix theory is presented in Section IV, where we also discuss
the range of validity of the correspondence between semiclassical and
statistical theories.  In Section V we present the conclusions of this
study.

\section{Wigner--Smith time delay and trapped periodic orbits}

This section is devoted to the derivation of a semiclassical equation
for the Wigner-Smith time delay in terms of classical periodic orbits
trapped inside the scattering region.  Our derivation relies on the
association of the $S$-matrix with the quantum Poincar\'e map,
following closely the formalism developed by Bogomolny
\cite{Bogomolny92} for closed systems. A similar result was previously
obtained by Balian and Bloch \cite{Balian74}, based on a construction
proposed by Friedel \cite{Friedel52} for a separable system.  The
derivation presented below is more transparent than the one in Ref.
\onlinecite{Balian74}, making the approximations more controllable when
dealing with actual systems.

We begin by showing a simple construction relating the energy
derivatives of two sets of invariants of the $\Lambda$-channel
scattering matrix $S$, namely the eigenphases $\{\theta_1,\theta_2,
\ldots, \theta_\Lambda\}$ and the traces $\{\mbox{Tr}\,
S^n,~n=1,2,\ldots\}$ (the construction remains valid for an arbitrary
unitary matrix depending on one parameter). For this purpose, let us
consider the periodic function
\begin{equation}
F(\theta)= \theta~\mbox{mod} \, 2\pi \;,
\end{equation}
which has a Fourier expansion given by 
\begin{equation}
\label{eq:ftheta}
F(\theta)= \pi -2 \sum_{n=1}^{\infty} \frac{\sin n\theta}{n} \; .
\end{equation}
Summing over all $S$-matrix eigenphases in both sides in
Eq.(\ref{eq:ftheta}) and using $\mbox{Im}\,\mbox{Tr}\, S^n =
\sum_{c=1}^{\Lambda} \sin n\theta_c$ we obtain
\begin{equation}
\label{eq:sftheta}
\sum_{c=1}^{\Lambda} F(\theta_c)= 
\Lambda \pi -2 \,\mbox{Im}\, \sum_{n=1}^{\infty} 
\frac{1}{n} \mbox{Tr}\, S^n \;.
\end{equation}
The convenience of this arbitrary choice of $F$ becomes apparent after
differentiating all terms in (\ref{eq:sftheta}) with respect to the
energy $E$,
\begin{equation}
\label{smilansky}
\sum_{c=1}^{\Lambda} \frac{\partial \theta_c}{\partial E}  = 
2 \pi \sum_{c=1}^{\Lambda} \frac{\partial \theta_c}{\partial E}
\, \delta( \theta_c ~\mbox{mod} \, 2\pi)- 
2\, \mbox{Im}\, \sum_{n=1}^{\infty} \frac{1}{n} \frac{\partial 
\mbox{Tr}\, S^n}
{\partial E} \;,
\end{equation}
since by recalling (\ref{twsdef}) it is easy to identify the l.h.s.~of
the above expression with the Wigner-Smith time delay.
Eq.~(\ref{smilansky}) was first obtained by Bogomolny\cite{Bogomolny92}
and later rederived by Rouvinez and Smilansky\cite{Rouvinez95}. Those
authors were interested in using the transfer (or scattering matrix)
approach to develop a quantization procedure for closed systems.  The
ideas presented below are quite different, since we are interested in
open systems.  Indeed, we use the closed system to understand the
scattering problem, which is the reverse of the procedure in
Refs.~\onlinecite{Rouvinez95,Doron92}.  To this end,
Eq.~(\ref{smilansky}) only becomes useful after the following steps.

First, we shall consider the scattering matrix $S$ (at the energy $E$)
for the specific system discussed in Section I.  Far away from the
cavity, at $x = 0$, the influence of the evanescent modes to the wave
functions is negligible, since $|k_c| D \gg 1$ (see Fig.~1).
Therefore, according to Eq.~(\ref{scatwave}), the exact quantization
condition for the system closed at $x=0$,  becomes $\det(S-1)=0$, as
has already been observed \cite{Doron92}.  In other words, one of the
eigenphases of the $S$-matrix must vanish
\begin{equation}
\label{quatization}
\theta_c~\mbox{mod} \, 2\pi = 0 \;.
\end{equation}
With this quantization condition, the first term in the r.h.s. of 
Eq.~(\ref{smilansky}) is now easily identified with the density of 
states, $\rho_{L+R}(E)$, of the system closed at $x=0$.
After proper averaging over some energy interval, $\rho_{L+R}(E)$
can be decomposed into a smooth and a fluctuating part
\begin{equation}
\sum_{c=1}^{\Lambda} \frac{\partial \theta_c}{\partial E}
\, \delta( \theta_c ~\mbox{mod} \, 2\pi)= 
\rho_{L+R}^{\mbox{\scriptsize{av}}}(E) +
\rho_{L+R}^{\mbox{\scriptsize{fl}}}(E) \;,
\label{relacao1}
\end{equation}
where $L$ and $R$ stand for the pipe and cavity regions respectively
(see Fig.~1). 

In the present context, the matrix $S$ is interpreted as the {\sl
quantum Poincar\'e map} of the closed system $L+R$ associated with the
section ${\cal{S}}$ ($x=0^+$ in Fig.~1).  The construction is quite
obvious, but for sake of completeness let us be explicit: Take an
asymptotic incoming wavefunction at $x=0^+$ and let it be scattered by
$R$.  As a result, one has a matrix that matches the incoming
asymptotic waves into the outgoing ones. This defines the quantum
return map for $x=0^+$ and it is also the definition of the
$S$-matrix.  In order to obtain the Poincar\'e map, one still needs to
describe what happens for $x < 0^-$.  This, however, is trivial, since
$x<0^-$ defines the asymptotic region (there is no coupling between
channels) and the corresponding Poincar\'e map is the identity matrix.
For the geometry considered here, the reflection by the hard wall is
equivalent to the reinjection procedure, defining the so called
Poincar\'e {\sl scattering} map, originally proposed by Jung
\cite{Jung97}.  In this way, the $S$-matrix can also be viewed as the
quantization of the Poincar\'e scattering map.
It is noteworthy that the quantization condition defined by
Eq.~(\ref{quatization}) has an interesting semiclassical counterpart.
For closed systems, the accuracy of the semiclassical quantization
procedure based on a Poincar\'e section requires the section to be
traversed by all periodic orbits, such a condition defining a ``good"
section.  If this is not the case, evanescent corrections become
essential.  By closing the system sufficiently far away from the cavity
region we ensure that the evanescent contributions die out, making the
exact quantum problem simpler.

Since the $S$ matrix can be obtained by the quantization of 
a classically chaotic map, in the semiclassical approximation
its traces can be expressed as a sum over periodic orbits.  
Actually, the sum over the traces of $S$ in Eq.~(\ref{smilansky}) 
results in an expression very similar to the standard Gutzwiller 
trace formula\cite{Gutzwiller90} for the oscillatory part of the 
density of states of the system defined by $L+R$\cite{Rouvinez95}.
The important difference is that in our case the sum is restricted 
to those periodic trajectories {\sl that touch the section 
${\cal{S}}$\/}.
Decomposing the full set of periodic orbits of the system $L+R$
into a set that reaches ${\cal{S}}$
and a set which never leaves the cavity $R$, one can write
\begin{equation}
\label{relacao4}
\frac{1}{\pi} \mbox{Im} \sum_{n=1}^{\infty} \frac{1}{n} 
\frac{\partial}{\partial E} \mbox{Tr} \, S^n
\approx 
\rho_{L+R}^{\mbox{\scriptsize{fl}}}(E) -
\rho_{  R}^{\mbox{\scriptsize{fl}}}(E) \;,
\end{equation}
where $\rho_{R}^{\mbox{\scriptsize{fl}}}(E)$ can be expressed
in terms of periodic orbits constricted to the region $R$.

Substituting the relations (\ref{relacao1}) and (\ref{relacao4}) 
into Eq.~(\ref{smilansky}), we obtain
\begin{equation}
\frac{\Lambda}{2 \pi \hbar} \tau(E) 
\approx
\rho_{L+R}^{\mbox{\scriptsize{av}}}(E) +
\rho_{L+R}^{\mbox{\scriptsize{fl}}}(E)  -
\left(\rho_{L+R}^{\mbox{\scriptsize{fl}}}(E) -
      \rho_{  R}^{\mbox{\scriptsize{fl}}}(E) \right) \;,
\end{equation} 
yielding 
\begin{equation}
\label{eqtau0}
\tau(E) \approx 
\frac{2 \pi \hbar}{\Lambda} \Big(
\rho_{L+R}^{\mbox{\scriptsize{av}}}(E) +
\rho_{  R}^{\mbox{\scriptsize{fl}}}(E)  \Big) ~.
\end{equation} 
This is already very close to the expression we are looking for.  The
problem is that $\tau$ is measured with respect to $x=0$ and we are
interested in the time that the particle spends in the cavity region,
{\sl i.e.}, the time delay with respect to $x=D$.  As a consequence, we
still have to translate the origin of coordinates to the entrance of
the cavity.  Under this operation, the $S$-matrix transforms as
\begin{equation}
S(x')= e^{-i{\bf k}D} S(x) e^{-i{\bf k}D} \;,
\end{equation} 
where $x'=x-D$ and
${\bf k}$ is a diagonal matrix having the $k_c$'s as elements. 
Taking into account that the time delay is additive with respect to
the product of unitary operators, namely,
\begin{equation}
\mbox{Tr} 
          \Big((S_1 S_2)^\dagger \frac{d}{dE}(S_1 S_2) \Big) =
\mbox{Tr} \Big( S_1^     \dagger \frac{d}{dE} S_1      \Big) +
\mbox{Tr} \Big( S_2^     \dagger \frac{d}{dE} S_2      \Big) 
\end{equation}
we arrive at 
\begin{equation}
\label{eqtau1}
\tau' = \tau -
\frac{2 \hbar D }{\Lambda} 
\frac{d}{dE} \sum_{c=1}^{\Lambda} k_c \;.
\end{equation}  
The second term in the r.h.s. of (\ref{eqtau1}) is a smooth function of
energy.  It is proportional to the density of states of the region
$L$ (the pipe in Fig.~1),
\begin{equation}
\label{eqtau2}
\frac{2 \hbar D }{\Lambda} \frac{d}{dE} \sum_{c=1}^{\Lambda}  k_c  =
\frac{2 D}{\Lambda} \sum_{c=1}^{\Lambda} \frac{1}{v_c}  =    
\frac{2 \pi \hbar}{\Lambda}  \rho_{L}^{\mbox{\scriptsize{av}}}(E) \;.
\end{equation} 
The first identity above says that under a spatial translation by $D$,
the time delay varies by the classical time it takes to travel to the
point displaced by $D$ and back, averaged over the channels.  The
second identity is obtained by using the Weyl formula for the quasi
one-dimensional density of states in a waveguide, expressed in terms of
the longitudinal velocities $v_c=\hbar k_c/m$.  Inserting
(\ref{eqtau1}) and (\ref{eqtau2}) into (\ref{eqtau0}), we obtain
\begin{equation}
\label{tauxD}
\tau' \approx 
\frac{2 \pi \hbar}{\Lambda} \Big(
\rho_{R}^{\mbox{\scriptsize{av}}}(E) +
\rho_{R}^{\mbox{\scriptsize{fl}}}(E)  \Big) ~.
\end{equation} 
This is the main result of this section.
The first term in the r.h.s.~of Eq.~(\ref{tauxD}) represents the mean
time spent in the cavity, 
$\langle\tau\rangle=2\pi\hbar/(\Lambda\Delta)=\tau_H/\Lambda$, in
agreement with Levinson's theorem (which holds irrespective of whether
the underlying classical dynamics is chaotic or not; see {\sl e.g.}
\onlinecite{Lewenkopf91}).  The second term is given by
\cite{Gutzwiller90}
\begin{equation}
\label{tausc}
\tau^{fl}(E) \approx \frac{1}{\Lambda}
                 \sum_{\nu m} T_{\nu}(E) A_{\nu m}(E)
        e^{im{s}_{\nu}(E)/\hbar-i\frac{\pi}{2}\mu_{\nu m}} \;,
\label{tauflu}
\end{equation}
where the sum runs over the primitive periodic orbits $\nu$ {\sl which
do not leave the cavity} and their $m$-th repetitions. As usual, one needs
from each periodic orbit $\nu$ its period $T_\nu$, action ${s}$, Maslov
index $\mu_\nu$, and the amplitudes $A_\nu$ given in terms of the
monodromy matrix $M_\nu$ \cite{Gutzwiller90},
\begin{equation}
A_{\nu m}=\frac{1}{\sqrt{|\det(M^m_{\nu}-1)|}} ~.
\end{equation}
We base all considerations that follow on Eq.(\ref{tausc}).  One has to
keep in mind that Eq.(\ref{tausc}) is a semiclassical result, sharing
most of the usual limitations of the standard trace formula for bound
systems.  It must be also emphasized that for the quantization of the
bound systems, one is free to seek a Poincar\'e map which includes all
the periodic orbits (in our system this is the Birkhoff bounce map).
Then, the r.h.s.~of Eq.~(\ref{relacao4}) is zero and the ``delay time"
for this map is just the smooth Weyl density of states as discussed by
Rouvinez and Smilansky \cite{Rouvinez95}.  The scattering map is
predetermined in our case.  This is a chaotic, singular, discontinuous
classical map because of its exclusion of the repeller, which accounts
for the fluctuations of the time delay. For this reason we cannot use
the time delay calculated directly from the semiclassical scattering
map, reworking the theory on the basis of the repeller.

\section{Universal correlations in the quantum time delay}

Here we apply the results of the last section to analyze the
fluctuations of the time delay as a function of energy.  We study the
crossover regime in which time reversal symmetry is lost due to the
presence of an external magnetic field $B$.  Our derivation extends
previous results by Bohigas and collaborators \cite{Bohigas95} (closed
systems at the crossover regime) and by Eckhardt \cite{Eckhardt93}
(open systems with preserved time-reversal symmetry).  Our presentation
closely follows the discussion in \cite{Bohigas95}.

A usual measure to characterize the time delay fluctuations is
the correlation function
\begin{equation}
\label{Ktau}
K_{\tau}(\varepsilon, B) = 
      \left\langle \tau^{fl}(E+\varepsilon,B)
                   \tau^{fl}(E,B) \right \rangle_{E} ~,
\end{equation}
where $\langle \ldots \rangle_E$ stands for energy average.  The
semiclassical approach to the calculation of this correlator begins by
inserting (\ref{tauflu}) in (\ref{Ktau}) (the dependence of all
quantities on $B$ will now remain implicit to simplify the notation):
\begin{eqnarray}
K_{\tau}(\varepsilon) &  = & 
    \frac{1}{\Lambda^2}
    \sum_{\nu \, m \, \nu' m'}
    \mbox{\Large{$\langle$}} 
    T_{\nu}(E+\varepsilon) T_{\nu'}(E)
    A_{\nu m}(E+\varepsilon) A^\ast_{\nu' m'}(E)   \\
                      &     &   \times
\exp \left[ im{s}_{\nu }(E+\varepsilon)/\hbar-i\pi \mu_{\nu m}/2
          -im'{s}_{\nu'}(E)/\hbar-i\pi\mu_{\nu' m'}/2       
         \right]
    \mbox{\Large{$\rangle$}}_E   ~.
\label{KofE}
\end{eqnarray}
Here the average is taken over an energy interval $\delta E$ which, to
be meaningful, must be large as compared with the quantum scale
$\Delta$ in order to include many resonances.  On the other hand, for
practical purposes, $\delta E$ must be small enough to allow for the
use of classical perturbation theory.  According to these
considerations, the most important effect of varying the energy is on
the actions, as they are measured in units of $\hbar$.  We write
\begin{equation}
{s}(E+\varepsilon) 
\approx {s}(E)+\varepsilon T \;, ~~
A(E+\varepsilon) \approx A(E)\;, ~~\mbox{and}~~
T(E+\varepsilon) \approx T(E)\;.
\end{equation}

Next, to evaluate $K_{\tau}(\varepsilon)$, we use the ``diagonal
approximation", neglecting contributions of pairs of orbits having
distinct actions, as they cancel out upon averaging over energy.  This
approximation is accurate for orbits with periods shorter than some
critical value, whose typical scale is the Heisenberg time $\tau_H$.
We shall come back to this point later.

For $B=0$ only two kinds of pairs of orbits survive: the pairs of
identical orbits and the pairs of time-reversed partners.  As the
magnetic field $B$ grows, the contribution to the correlation function
of time-reversed partners gradually decreases.  Keeping both
contributions, we have
\begin{equation}
K_{\tau}(\varepsilon) =
           \frac{1}{\Lambda^2} 
           \sum_{\nu m}
           \left\langle 
           T^2_{\nu} \, |A_{\nu m}|^2 
           e^{i \varepsilon m T_{\nu}/\hbar} 
           \left[ 1 + e^{i m \delta {s}_{\nu}/\hbar}  \right] 
           \right\rangle_E \;,
\end{equation}
where $\delta s \equiv \delta s(E, B)$ stands for the action 
difference between a pair of time-reversed orbits. 
Next, we group orbits having periods in a small interval $\delta t$ 
around $t$ (containing many orbits).
This introduces a kind of averaging procedure, defining smooth functions 
of $t$. 
We then integrate over all $t$ 
\begin{equation}
K_{\tau}(\varepsilon) =
           \frac{1}{\Lambda^2} 
           \int_{-\infty}^{\infty} dt \,          
           |t|\, e^{i \varepsilon t/\hbar} 
           \Bigg\langle 
           \, \left\langle 
           |t|\,A^2 
           \right\rangle_t
           \left\langle 
           \left[ 1 + e^{i \delta {s}/\hbar}     \right] 
           \right\rangle_t \,
           \Bigg\rangle_E  \;,
\end{equation}
where we discarded the multiple repetitions $|m| \ne 1$, which are
exponentially negligible with respect to primitive orbits.

To evaluate the time average of the amplitudes $A$ we use the sum rule 
for open chaotic systems \cite{Kadanoff84,Cvitanovic91}:
\begin{equation}
\label{opensumrule}
           \left\langle 
           |t|\,A^2 
           \right\rangle_t = e^{-\gamma(E) |t|} \;,
\end{equation}
where $\gamma(E)$, the so called escape rate, is $\gamma =
1/\tau_{\mbox{\scriptsize dwell}}$, with $\tau_{\mbox{\scriptsize
dwell}}$ defined in the Section I.  Eq.~(\ref{opensumrule}) has a
simple physical interpretation:  As we increase the period, periodic
orbits proliferate exponentially (with a rate given by the entropy) as
their stability tend to decrease also exponentially (the rate given by
the sum of positive Lyapunov exponents). For $\gamma = 0$, both effects
cancel and one recovers the sum rule based on the uniformity principle
for closed systems\cite{Hannay84}. For open systems the disbalance
between the entropy and Lyapunov exponents is just the escape rate
resulting in (\ref{opensumrule}) \cite{Eckmann85,Gaspard89}.

The time average of the crossover term has been discussed in detail in
Ref.~\cite{Bohigas95}.  Further theoretical arguments \cite{Berry86}
and numerical evidence \cite{Bruus96} suggest the exponential decay
\begin{equation}
            \left\langle 
           1 + e^{i \delta s/\hbar}    
           \right\rangle_t = 1+ e^{-B^2 \kappa(E)|t|/\hbar^2} ~.
\end{equation}
Here $\kappa(E)$ is a purely classical quantity, which measures the
rate of decay of the appropriate correlations in the chaotic system.
For more details see Ref.~\onlinecite{Bohigas95}.
Identifying $\gamma(E)$ and $\kappa(E)$ with their mean values (the
energy averaging interval is small) we write
\begin{equation}
K_{\tau}(\varepsilon) = \int_{-\infty}^{\infty} 
        dt \, |t| \, e^{i \varepsilon t/\hbar - \gamma |t|} 
        \left(1+ e^{-B^2 \kappa(E)|t|/\hbar^2} \right) \;.
\end{equation}
This integral is easily evaluated, resulting in
\begin{equation}
\label{Kofw}
K_{\tau}(\omega) = \frac{1}{2} \langle \tau \rangle^2
\left\{ \frac{\Gamma^2-\omega^2}{[\Gamma^2+\omega^2]^2} + 
        \frac{(\Gamma + y)^2-\omega^2}{[(\Gamma + y)^2+\omega^2]^2}
\right\} ~,
\end{equation}
where, for the sake of future comparisons, we have defined
\begin{equation}
\omega = \pi \varepsilon  / \Delta                     \;,~~ 
\Gamma = \pi \gamma \hbar / \Delta                     \;,~~
     y = \frac{B^2 \kappa \pi}{\hbar \, \Delta}        \;.
\end{equation}
When $B=0$, we note that Eq.~(\ref{Kofw}) reduces to the result
obtained by Eckhardt\cite{Eckhardt93} for the time reversal symmetric
case. Alternatively, putting $\gamma = 0$ we recover the results of
Ref.\onlinecite{Bohigas95} for the density--density correlator of
the closed problem. 

One of the most interesting properties of the semiclassical
approximation to $K_\tau(\omega)$ is that in this case we obtain an
intrinsically more accurate result than the density-density correlators
for closed systems, that have been extensively semiclassically studied.
The reason is that the semiclassical approximation starts to fail for
energy domains of the order of the mean level spacing $\Delta$,
corresponding to times longer than $\tau_{\mbox{\scriptsize H}} =
2\pi\hbar/\Delta$, and such times are normally unimportant for the computation of $K_\tau$.  
The physics of the scattering problem provides us with a
natural cut-off for the summation over periodic orbits, which is given
by the typical escape time $1/\gamma$, explicit in
(\ref{opensumrule}).  Since the semiclassical theory is only applicable
if the waveguide $L$ has many open channels, $1/\gamma$ must be
much smaller than the Heisenberg time $\tau_H$, corresponding to the
regime of strongly overlapping resonances.  Although we do not have a
control over the magnitude of the accuracy, by increasing the number of
open channels in actual systems, $K_\tau(\omega)$ converges to the
semiclassical approximation (\ref{Kofw}).

\section{Comparison with the statistical approach}

The statistical properties of the Wigner-Smith time delay for chaotic
systems have been also investigated by using the Random Matrix Theory
\cite{Harney92,Dittes92,Muga95,Lehmann95a,Fyodorov97b}.  In analogy
with the semiclassical approach, the statistical approach requires a
decomposition of the Hilbert space into an asymptotic region $L$ and a
``complex" scattering region $R$, following the notation of Fig.~1.  As
before, depending on the energy $E$, one will have $\Lambda$
propagating channels in the pipe, labelled by $c$.  By introducing
arbitrary boundary conditions at $x=D$ one can define a set of
quasi-bound states $|\mu\rangle$ in region $R$ and a set of scattering
states $|\chi_c(E)\rangle$ in $L$.  These states form a complete set $Q
+ P = 1$, with $Q = \sum_\mu |\mu\rangle \langle \mu|$  and $P = \sum_c
\int dE |\chi_c(E)\rangle \langle\chi_c(E)|$.  Thus, the Hamiltonian is
given by
\begin{equation}
H = QHQ + PHP + QHP + PHQ \equiv H_{QQ} + H_{PP} + H_{PQ} + H_{QP} \;,
\end{equation}
where $H_{QQ}$ is interpreted as an ``internal" Hamiltonian, and $H_{PQ}$ 
($H_{QP}$) are the couplings between the interior $R$ and the channel 
region $L$ .
After some algebra, one can show \cite{Mahaux69,Lewenkopf91} that the
resonant $S$-matrix can be written as
\begin{equation}
\label{Smatrix}
S = I - 2 \pi i H_{PQ} \frac{1}{E - H_{QQ} + i\pi H_{QP}\, H_{PQ}}
                    H_{QP} \;,
\end{equation}
in the absence of direct reactions, implying that $H_{PP}$ is
diagonal.  The decomposition of the Hilbert space in projectors $P$ and
$Q$ can, in principle, be employed for a large variety of problems,
which makes Eq.~(\ref{Smatrix}) a very useful parameterization of the
$S$-matrix.  It follows that the Wigner-Smith time delay is given by
\cite{Fyodorov97b}
\begin{equation}
\label{tauPQ}
\tau(E) = -\frac{2}{\Lambda} \mbox{Im} \, \mbox{Tr} \,
              \Big( E - H_{QQ} + i\pi H_{QP}\, H_{PQ}\Big)^{-1}  \;.
\end{equation}
This expression is akin to the one obtained semiclassically
(\ref{tauxD}), as it should.  Here $\tau(E)$ is equated to the level
density of the ``closed" system, defined by the operator $Q$ (which is
quite arbitrary) and smoothed by the coupling to the exterior world by
the imaginary term in (\ref{tauPQ}).  Although conceptually similar to
Eq.~(\ref{tauxD}), it is not a simple task to arrive at the 
semiclassical expression starting from Eq.~(\ref{tauPQ}).

Since one expects signatures of chaos in scattering processes to be
manifest for times much longer than the typical traversal time, we
focus our attention only in the resonant part of $S$.  This is the
physical justification for neglecting direct (fast) reactions in
Eq.~(\ref{Smatrix}).  Moreover, the object which is responsible for
classical chaos in scattering is the repeller, implying that chaos is a
property of the ``internal" Hamiltonian.  Therefore, in analogy with
Bohigas' conjecture \cite{Bohigas84} for closed systems, a statistical
modelling of quantum chaos in open systems can be made by taking
$H_{QQ}$ as a member of an ensemble of random matrices
\cite{Lewenkopf91}.  For instance, for preserved time-reversal symmetry
$H_{QQ}$ belongs to the Gaussian Orthogonal Ensemble (GOE) and for
broken time-reversal symmetry to the Gaussian Unitary Ensemble (GUE).
This conjecture allows us to study universal fluctuations in scattering
processes by calculating $S$-matrix correlation functions.  Those are
obtained by ensemble averaging, which is equivalent to an energy
averaging based on the ergodic hypothesis.

In particular, the calculation of the 2-point time delay correlation
function $K_\tau(\varepsilon, Y)$, studied in the previous section,
requires the evaluation of
\begin{equation}
\label{Krmt}
K_\tau(\varepsilon, Y) = \frac{2}{\Lambda^2} \mbox{Re} \left\{
   \overline{\mbox{Tr}\,g        (E + \frac{\varepsilon}{2},Y)
             \mbox{Tr}\,g^\dagger(E - \frac{\varepsilon}{2},Y)} -
   \overline{\mbox{Tr}\,g        (E,                        Y)}^2
\right\}    \;,
\end{equation}
where $\overline{\cal{O}}$ denotes the ensemble average of ${\cal{O}}$ 
and
\begin{equation}
g(E, Y) = \Big( E - H_{QQ}(Y) + i \pi H_{QP} H_{PQ}\Big)^{-1} \;,
\end{equation}
with the variable $Y$ parameterizing changes in the internal
Hamiltonian.  For instance, if $Y$ stands for an external magnetic
field $B$, one can study $K_\tau(\varepsilon, Y)$ in the crossover
regime between preserved and broken time reversal symmetry by choosing
$H_{QQ}=H^{GOE}+ Y H^{GUE}$.  The results are universal in terms of the
scaled variable $y = Y/Y_c$, where $Y_c$ is system specific.

From the technical point of view, the ensemble average in
Eq.~(\ref{Krmt})
implies a nontrivial calculation based on the supersymmetric technique
developed by Efetov \cite{Efetov83}.  This technique was adapted to
scattering problems by the Heidelberg group
\cite{Verbaarschot85,Verbaarschot86}.  Ref.~\onlinecite{Verbaarschot85}
is the starting point of all works that use the statistical approach to
study the time delay correlation function
\cite{Lehmann95a,Fyodorov97a,Fyodorov97b} and related objects
\cite{relatedstuff}. A discussion of the supersymmetric technique is
beyond the scope of this paper and we refer the reader to the excellent
introductory text in Ref.~\onlinecite{Fyodorov95}, and to the $S$-matrix
review in Ref.~\onlinecite{Fyodorov97b}.

Let us start with the simplest case, $K_\tau(\omega, y\gg 1)$ for
broken time-reversal symmetry, corresponding to taking $H_{QQ}$ as a
member of the Gaussian Unitary Ensemble. The result, given as usual in
terms of a double integral, can be found in
Ref.\onlinecite{Fyodorov97b}
\begin{equation}
\label{KGUE1}
K_\tau^{GUE}(\omega) = \frac{\langle\tau\rangle^2}{2}
\int_{-1}^{1}d\lambda
\int_1^\infty d\lambda_1 \cos\Big(\omega(\lambda_1 - \lambda)\Big)
\prod_{c=1}^\Lambda \left(\frac{2 + T_c(\lambda - 1)}{2 + T_c(\lambda_1-1)}
\right)
\end{equation}
where the transmission coefficient $T_c$ gives the probability of
an incoming wave at the channel $c$ in the vicinity of $x=D$ to enter 
the scattering region $R$.
In order to compare with the semiclassical theory, one has to
take $T_c = 1$ for all channels $c$, since this theory does not 
take into account any barriers preventing perfect transmission. 
Thus, keeping the notation introduced in Section III and
identifying $\langle\tau\rangle=1/\gamma$, the leading 
asymptotic term in powers of $\Gamma^{-1}$ of Eq.~(\ref{KGUE1}) 
becomes
\begin{equation}
K_\tau^{GUE}(\omega) \approx 
\frac{\langle\tau\rangle^2}{2} \; 
     \frac{\Gamma^2 - \omega^2} {[\Gamma^2 + \omega^2]^2} \;,
\end{equation}
in nice agreement with the semiclassical result.  
%
%
In Fig.~2 we present $K_\tau^{GUE}$ as a function of the number of open
channels $\Lambda$.  As $\Lambda$ increases, the agreement with the
semiclassical theory becomes much better, as it is nicely shown in the
inset of Fig.~2.  Even for relatively small $\Lambda$, the exact result
does not differ significantly from the semiclassical one.  This is
explained by the fact that one can show that the next to leading order
correction is smaller by a factor $\Gamma^{2}$.

The other simple limit is the case where time reversal symmetry is
present, corresponding to $H_{QQ}$ taken as a member of the GOE.
Here the result is \cite{Lehmann95a}
\begin{eqnarray}
K_\tau^{GOE}(\omega) & = & \frac{\langle\tau\rangle^2}{4}
  \int_0^1 \! d\lambda \int_0^\infty\! d\lambda_1 \int_0^\infty \!d\lambda_2 
  \,\mu(\lambda, \lambda_1, \lambda_2) 
  (2\lambda + \lambda_1 + \lambda_2)^2   \\     
  &  & \times \cos\Big(\omega (2\lambda + \lambda_1 + \lambda_2) \Big) 
       \prod_c \left( \frac{(1 - T_c \lambda )^2}
       {(1 + T_c \lambda_1)(1 + T_c \lambda_2)} \right)^{1/2} ~,
\end{eqnarray}
with the measure $\mu$ given by
\begin{equation}
\mu(\lambda, \lambda_1, \lambda_2) =
   \frac{(1 - \lambda)\lambda |\lambda_1  - \lambda_2|}
    {[(1+\lambda_1)\lambda_1 (1 + \lambda_2) \lambda_2]^{1/2}
     (\lambda + \lambda_1)^2(\lambda + \lambda_2)^2} \;.
\end{equation}
In this case, even taking $T_c = 1$ for all $c$'s the integral is still
difficult. However, we can use a trick introduced by Efetov
\cite{Efetov83} or the asymptotic expansion proposed by Verbaarschot
\cite{Verbaarschot86} to obtain
\begin{equation}
K_\tau^{GOE}(\omega) \approx 
\langle\tau\rangle^2 \; 
     \frac{\Gamma^2 - \omega^2} {[\Gamma^2 + \omega^2]^2} \;,
\end{equation}
again in nice agreement with the semiclassical result. Here, higher
order corrections are smaller only by a factor $\Gamma$. 
This is manifest in
Fig.~3, where one can see that the GUE case converges faster than
the GOE to the semiclassical result.

For the crossover regime the supersymmetric expressions become
even more complicated. By performing a calculation similar to
the one done by Pluha\v{r} and collaborators \cite{Pluhar95}, 
Fyodorov, Savin, and Sommers \cite{Fyodorov97a} obtained
a closed expression for $K_\tau$, first numerically studied
in Ref.\onlinecite{Jalabert97} as the ballistic limit of
electronic mesoscopic transmission.
The leading asymptotic term in inverse powers of $\Gamma$ is 
\begin{equation}
K_\tau (\omega, y) = \frac{\langle\tau\rangle^2}{2} \left( 
     \frac{\Gamma^2 - \omega^2} {[\Gamma^2 + \omega^2]^2} +
     \frac{(\Gamma+y)^2 - \omega^2} {[(\Gamma+y)^2 + \omega^2]^2} 
\right) \;,
\end{equation}
identical to the semiclassical result Eq.~(\ref{Kofw}).
 
The strength of the statistical method is in dealing with situations
beyond the scope of the semiclassical theory, either by analyzing
situations where $\Lambda$ is small, or by treating systems where the
scattering waves have to overcome barriers.  In the latter case, the
semiclassical analysis needs some refinements.  Equally or even more
problematic is the fact that when $\sum T_c < 1$, the resonances become
isolated. This calls for a semiclassical theory
that deals with times larger than $\tau_H$.

This is a good point to call the attention to one
issue in the literature\cite{Gaspard89} that is often confusing.
The ensemble average is
performed for a fixed number of channels and many resonances between
channel thresholds. The accuracy of the statistical approach increases
as the number of resonances increase.  With such a construction it is not
possible to use this theory to statistically analyze the $S$-matrix of
a system like the 3-disc problem, since between channel thresholds
there are very few resonances, explaining the disagreement in
\onlinecite{Gaspard89}. However, since one takes the trace over the
channels to obtain the Wigner-Smith time delay, provided $\Lambda \gg
1$, a small change in $\Lambda$ seems not to affect $K_\tau$, as
observed in Ref.~\onlinecite{Eckhardt93}.

\section{Conclusions}

In this paper we presented a semiclassical derivation of the formula
connecting the Wigner-Smith time delay $\tau$ to the resonance density
of the scattering region, corresponding to a chaotic system.  
We showed that $\tau$ can be written as a sum over the periodic orbits 
inside the repeller. 
The physical interpretation of this relation is that the repeller
is responsible for the time spent in the cavity by the scattering
trajectories. 
An open trajectory that closely approximates a periodic one, can spend a
long time in the scattering region. This dwell time essentially depends 
on the stability 
of the periodic trajectory that is approached.
As a result, the typical classical dwell time depends on few bulk 
characteristics of the scattering system.
The interesting achievement of the semiclassical theory is that 
it is possible to write scattering observables in terms of classical
trajectories that never leave the system.
Doing so, one avoids all problems inherent of a semiclassical formulation 
in terms of open trajectories.

One of the striking features of the semiclassical approximation for a
scattering system is its accuracy.
In distinction to the usual studies of density-density correlators in
closed systems, here one has no need to account for times of the order
of $\tau_H$, the time scale where the semiclassical approach starts
failing.
Trajectories entering an open chaotic system cannot typically  stay
inside the scattering region for times much longer than $\langle \tau 
\rangle$.
This fact provides us with a natural cut-off for any semiclassical
summation formula, namely $\langle \tau \rangle$ itself. 
If we assume that convergence is the only problem of the semiclassical
formalism, systems with increasing numbers of open channels will be 
described by the semiclassical theory with increasing precision as 
compared with the exact theory.

Although we already know the exact statistical result for several
correlators and distributions involving the Wigner-Smith time delay,
such an approach, by construction, does not have information about 
non-universal quantities (like $\Delta$, $Y_c$, etc..).
Those are usually extracted from the experiment.
The point of this paper is that this information is usually 
available from the classical dynamics and the semiclassical approach 
can always be adapted to give a recipe to compute the non-universal 
scaling factors.
In summary, even if the semiclassical theory cannot compete in accuracy, 
it can be used as a complement to the statistical approach.

\acknowledgements

We would like to thank Eduardo R. Mucciolo for many interesting 
discussions.
This work was supported by the Conselho de Desenvolvimento 
Cient\'{\i}fico e Tecnol\'ogico (CNPq/Brazil), by the 
Centro Latino Americano de F\'{\i}sica (CLAF), and by the
Funda\c{c}\~ao de Amparo \`a Pesquisa do Rio de Janeiro 
(FAPERJ/Brazil).



\begin{figure}
\caption{Schematic illustration of the scattering system under
investigation. The ``cavity" region is denoted by $R$ and the attached
waveguide by $L$.}
\label{FIG1}
\end{figure}

\begin{figure}
\caption{Comparison between the semiclassical (solid lines) and the
exact random matrix results for the time--delay correlator (case where time-reversal symmetry is absent). We plot the normalized correlators
$K_\tau^\ast=2\Gamma^2 K_\tau/\langle\tau\rangle^2$ vs. 
the normalized energy $\omega/\Gamma$. 
Different line styles indicate different number of open channels; 
dashed, dash--dot, and dotted correspond to $\Lambda=3,5,10$,
respectively. Inset: difference between random matrix results and
semiclassics.}
\label{FIG2}
\end{figure}

\begin{figure}
\caption{Comparison between the semiclassical (solid lines) and the
exact random matrix results for the time--delay correlator (case of preserved time-reversal symmetry). We plot the normalized correlators
$K_\tau^\ast=\Gamma^2 K_\tau/\langle\tau\rangle^2$ vs. 
the normalized energy $\omega/\Gamma$. 
Different line styles indicate different number of open channels; 
dashed, dash--dot, and dotted correspond to $\Lambda=3,5,10$,
respectively. Inset: difference between random matrix results and
semiclassics.}
\label{FIG3}
\end{figure}

\newpage

\epsfxsize=10.0cm
\epsfbox[223 213 399 550]{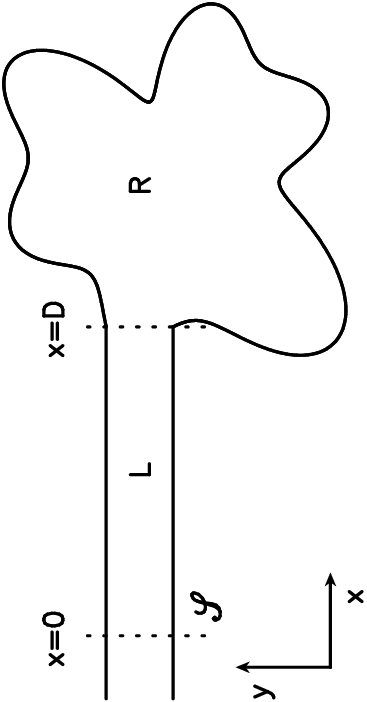}    

\newpage

\epsfxsize=15.0cm
\epsfbox[129 100 525 631]{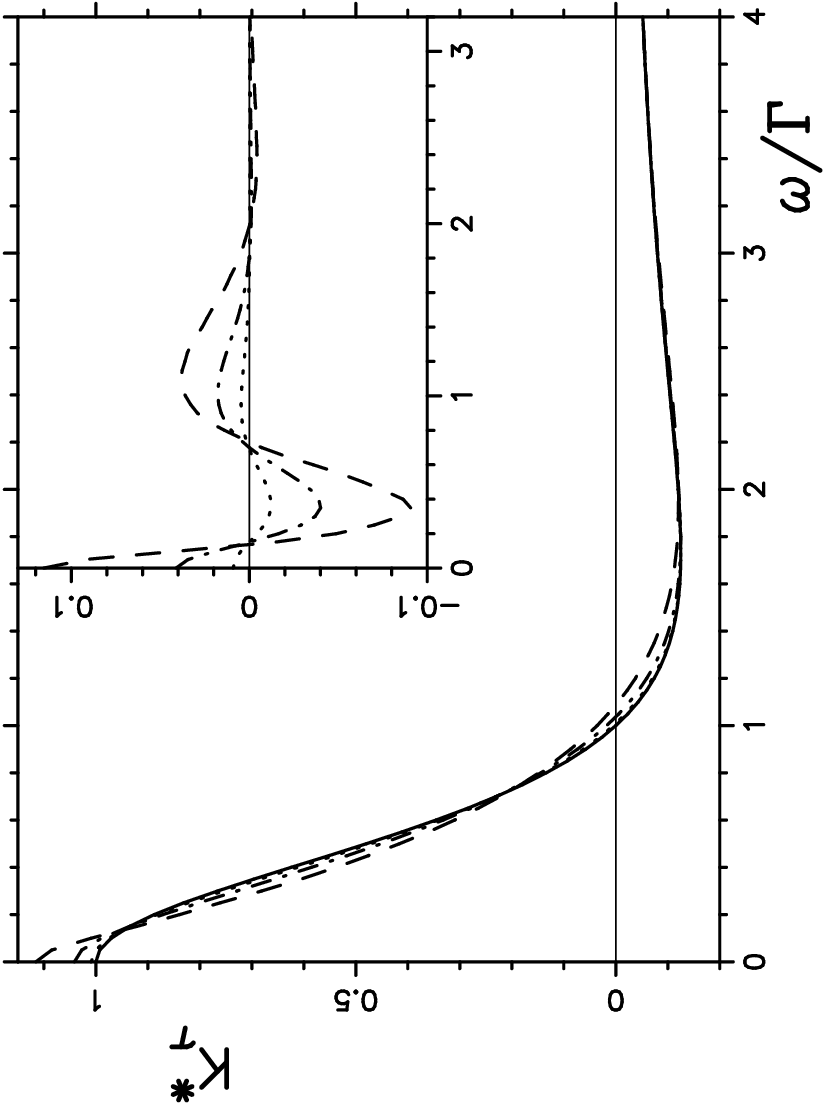}    

\newpage

\epsfxsize=15.0cm
\epsfbox[129 100 525 631]{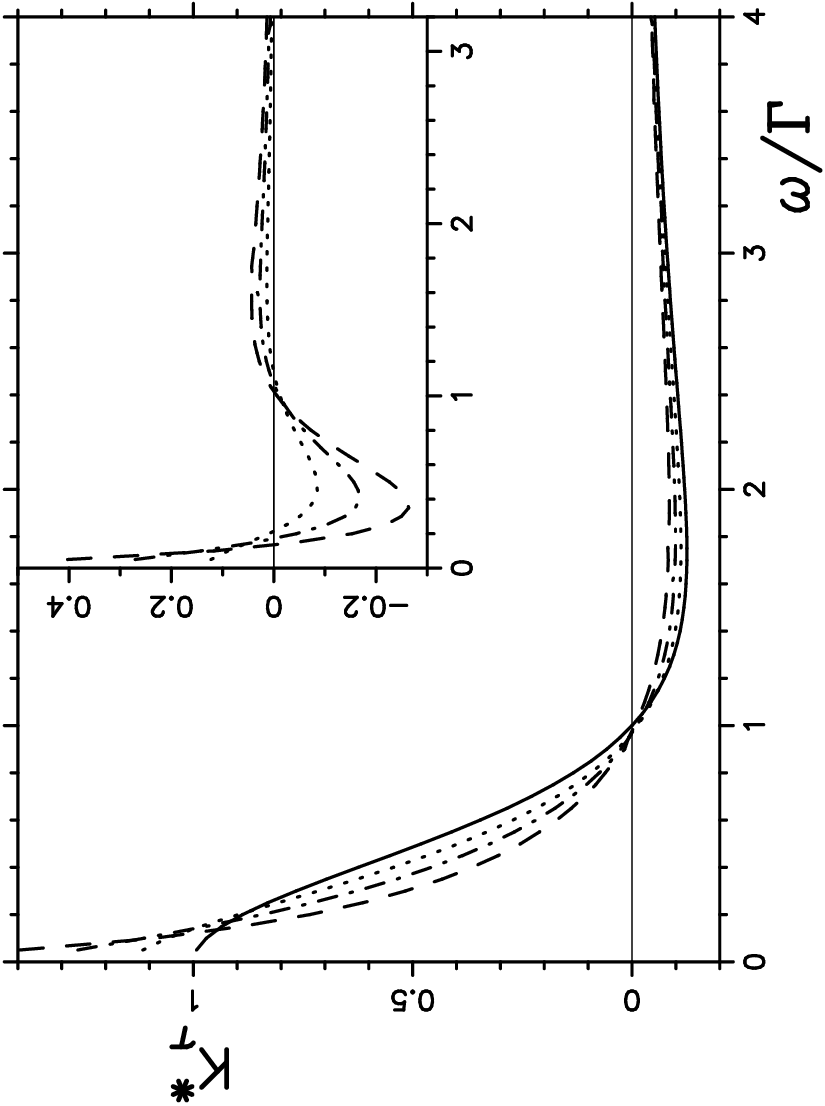}


\begin{thebibliography}{99}

\bibitem{Wigner55}
    E. P. Wigner,
         Phys. Rev. {\bf 98}, 145 (1955).

\bibitem{Smith60} 
    F. T. Smith,
         Phys. Rev. {\bf 118}, 349 (1960).

\bibitem{Eisenbud48}
    L. E. Eisenbud, 
         {\sl PhD Thesis}, Princeton University, 1948 (unpublished).

\bibitem{Nussenszveig72}
    H. M. Nussenszveig,
         Phys. Rev. D. {\bf 6}, 1534 (1972); 
         Phys. Rev. A. {\bf 55}, 1012 (1997). 

\bibitem{Buttiker93}
    M. B\"uttiker, A. Pr\^etre, and H. Thomas, 
         Phys. Rev. Lett. {\bf 70}, 4114 (1993).

\bibitem{others}
    V. A. Gopar, P. A. Mello, and M. B\"uttiker, 
         Phys. Rev. Lett. {\bf 77}, 3005 (1996); 
    P. W. Brouwer and M. B\"uttiker, 
         Europhys. Lett. {\bf 37}, 441 (1997); 
    P. W. Brouwer, S. A. van Langen, K. M. Frahm, M. B\"uttiker,
    and C. W. Beenakker,
         Phys. Rev. Lett. {\bf 79}, 913 (1997).

\bibitem{Doron90}
    E. Doron, U. Smilansky, and A. Frenkel, 
         Phys. Rev. Lett. {\bf 65}, 3072 (1990).

\bibitem{Lewenkopf92}
    C. H. Lewenkopf, A. M\"uller, and E. Doron,
         Phys. Rev. A {\bf 45}, 2635 (1992).

\bibitem{Alt95}
   H. Alt, 
    H.--D.~Gr\"af, H.~L.~Harney, R.~Hofferbert, H.~Lengerer,
    A.~Richter, P.~Schardt, and H.~A.~Weidenm\"uller,
                 Phys. Rev. Lett. {\bf 74}, 62 (1995).

\bibitem{Bogomolny92}
   E. Bogomolny, 
         Nonlinearity {\bf 5}, 805 (1992).

\bibitem{Balian74}
    R. Balian and C. Bloch, 
         Ann. Phys. {\bf 85}, 514 (1974).

\bibitem{Friedel52}
    J. Friedel,
         Phylos. Mag. {\bf 43}, 153 (1952).

\bibitem{Rouvinez95}
   C. Rouvinez and U. Smilansky, 
        J. Phys. A: Math. Gen. {\bf 28}, 77 (1995).
     
\bibitem{Doron92}
   E. Doron and U. Smilansky,
         Nonlinearity {\bf 5}, 1055 (1992).

\bibitem{Jung97}
   C. Jung and T. H. Seligman,
         Phys. Rep. {\bf 285}, 77 (1997). 

\bibitem{Gutzwiller90}
   M. C. Gutzwiller, 
        {\sl Chaos in Classical and Quantum Mechanics}
         (Springer, New York, 1990).

\bibitem{Lewenkopf91}
   C. H. Lewenkopf and H. A. Weidenm\"uller, 
         Ann. Phys. {\bf 212}, 53 (1991).  

\bibitem{Bohigas95}
    O. Bohigas, M.--J. Giannoni, A. M. O. de Almeida, and C. Schmit,
         Nonlinearity {\bf 8}, 203 (1995).

\bibitem{Eckhardt93}
   B. Eckhardt, 
         Chaos {\bf 3}, 613 (1993).

\bibitem{Kadanoff84}
    L. P. Kadanoff and C. Tang, 
        Proc. Natl. Acad. Sci. USA {\bf 81}, 1276 (1984).

\bibitem{Cvitanovic91}
   P. Cvitanovi\'c and B. Eckhardt, 
        J. Phys. A: Math. Gen. {\bf 24}, L237 (1984).

\bibitem{Hannay84} 
    J. H. Hannay and A. M. Ozorio de Almeida, 
	J.\ Phys.\ A:  Math. Gen. {\bf 17}, 3429 (1984).

\bibitem{Eckmann85}
    J.--P. Eckmann and D. Ruelle, 
         Rev. Mod. Phys. {\bf 57}, 617 (1985).

\bibitem{Gaspard89}
    P. Gaspard and S. A. Rice,
      J. Chem. Phys. {\bf 90}, 2242, 2255 (1989).

\bibitem{Berry86}
    M. V. Berry and M. Robnik,
         J. Phys. A: Math. Gen. {\bf 19}, 649 (1986).

\bibitem{Bruus96} 
   H. Bruus, C.~H. Lewenkopf, and E.~R. Mucciolo,
	Phys. Rev. B {\bf 53}, 9968 (1996).

\bibitem{Harney92}
   H. L. Harney, F.~M. Dittes, and A. M\"uller,
        Ann. Phys. {\bf 220}, 159 (1992).

\bibitem{Dittes92}
   F.--M. Dittes, H. L. Harney, and A. M\"uller, 
        Phys. Rev A {\bf 45}, 701 (1992).

\bibitem{Muga95}
   J. G. Muga and D. M. Wardlaw,
        Phys. Rev. E {\bf 51}, 5377 (1995).

\bibitem{Lehmann95a}
    N. Lehmann, D. Savin, V. Sokolov, and H.--J. Sommers,
         Physica {\bf 86D}, 572 (1995).

\bibitem{Fyodorov97b}
    Y. V. Fyodorov and H.--J. Sommers,
         J. Math. Phys. {\bf 38},  1918 (1997).

\bibitem{Mahaux69}
    C. Mahaux and H. A. Weidenm\"uller, 
        {\sl Shell--Model Approach to Nuclear Reactions}
        (North--Holland, Amsterdam, 1969).        

\bibitem{Bohigas84}
    O.~Bohigas, M.--J.~Giannoni, and C.~Schmit, 
        Phys. Rev. Lett. {\bf 52}, 1 (1984).

\bibitem{Efetov83}
    K. B. Efetov,
        Adv. Phys. {\bf 32}, 367 (1983).

\bibitem{Verbaarschot85}
    J. J. M. Verbaarschot, H. A. Weidenm\"uller, and M. R. Zirnbauer,
        Phys. Rep. {\bf 129}, 367 (1985).

\bibitem{Verbaarschot86}
    J. J. M. Verbaarschot, 
        Ann. Phys. (N.Y.) {\bf 168}, 368 (1986).

\bibitem{Fyodorov97a}
    Y. V. Fyodorov, D. V. Savin, and H.--J. Sommers,
        Phys. Rev. E {\bf 55}, R4857 (1997).

\bibitem{relatedstuff}
   V. V. Sokolov and V. G. Zelevinsky, 
         Nucl. Phys. {\bf A504}, 562 (1989); 
         Ann. Phys. {\bf 216}, 323 (1992),
   N. Lehmann, D. Savin, V. Sokolov, and H.--J. Sommers,
         Nucl. Phys. {\bf A582}, 223 (1995);
   Y. V. Fyodorov and H.--J. Sommers,
        Phys. Rev. Lett. {\bf 76}, 4709 (1996).

\bibitem{Fyodorov95}
    Y. V. Fyodorov, 
        in {\sl Mesoscopic Quantum Physics}, E. Akkermans,
        G. Montambaux, J.--L. Pichard, and J. Zinn--Justin, Eds.
        (Elsevier, Amsterdam, 1995).

\bibitem{Pluhar95}
    Z. Pluha\v{r}, 
        H. A. Weidenm\"uller, J. Zuk, C. H. Lewenkopf, and F. Wegner,
        Ann. Phys. (N.Y.) {\bf 243}, 1 (1995)

\bibitem{Jalabert97}
    R. A. Jalabert, E. R. Mucciolo, and J.--L. Pichard, 
        J. de Physique I (in press).

\end{thebibliography}
\end{document}